\mag\magstep 1
\documentclass[graybox]{svmult}
\evensidemargin=\oddsidemargin
\usepackage{mathptmx}       
\usepackage{helvet} 
\usepackage{courier}      
\usepackage{type1cm}    
\usepackage{makeidx}  
\usepackage{graphicx}    
\usepackage{multicol}   
\usepackage[bottom]{footmisc}
\makeindex    
\usepackage[nospace]{cite}
\usepackage{amssymb}
\usepackage{amsmath}
\usepackage{url}
\usepackage{doi}

\usepackage{color,bm}
\date{}

  \newcommand{\beq}{\begin{equation}}
  \newcommand{\eeq}{\end{equation}}
  \newcommand{\beql}[1]{\begin{equation}\label{eq:#1}}

  \newcommand{\beqa}{\begin{eqnarray}}
  \newcommand{\eeqa}{\end{eqnarray}}
  \newcommand{\beqas}{\begin{eqnarray*}}
  \newcommand{\eeqas}{\end{eqnarray*}}
 \newcommand{\bA}{{\bf A}}
  \newcommand{\bP}{{\bf P}}
  \newcommand{\bS}{{\bf S}}
  \newcommand{\cH}{{\mathcal H}}
  \newcommand{\cI}{{\mathcal I}}
 \newcommand{\cK}{{\mathcal K}}
  \newcommand{\cL}{{\mathcal L}}
  \newcommand{\cO}{{\mathcal O}}
  \newcommand{\cS}{{\mathcal S}}
  \newcommand{\al}{\alpha}

  \newcommand{\da}{\dagger}

  \newcommand{\et}{\eta}

  \newcommand{\mb}{\mbox}
  \newcommand{\nn}{\nonumber}
  
  \newcommand{\ph}{\phi}
 \def\ps{\psi}
  \newcommand{\rh}{\rho}
  \newcommand{\si}{\sigma}
  \newcommand{\ta}{\tau}
  \newcommand{\tc}{\cL}

 \newcommand{\De}{\Delta}
  \newcommand{\Eq}[1]{Eq.~(\ref{eq:#1})}
 \newcommand{\Tr}{\mbox{\rm Tr}}

  \newcommand{\beqan}{\begin{eqnarray*}}
  \newcommand{\beqar}[1]{\begin{equation}\label{#1}\begin{array}{l}}
  \newcommand{\bx}{{\bf x}}
  \newcommand{\by}{{\bf y}}
   
  \newcommand{\eq}[1]{(\ref{eq:#1})}
  \newtheorem{Theorem}{Theorem}
  
 \newcommand{\R}{\mathbb{R}}
\newcommand{\bra}[1]{\left\langle#1\right|}
\newcommand{\ket}[1]{\left|#1\right\rangle}
\newcommand{\ketbra}[1]{\ket{#1}\!\!\bra{#1}}
\newcommand{\bracket}[1]{\left\langle#1\right\rangle}
\newcommand{\id}{{\rm id}}

\begin{document}
\title*{Quantum Measurement Theory for\\ Systems with Finite Dimensional State\\ Spaces}
\titlerunning{Quantum Measurement Theory}
\author{Masanao Ozawa}
\institute{Masanao Ozawa \at Center for Mathematical Science and Artificial Intelligence, 
Chubu University Academy of Emerging Sciences, Chubu University, 
1200 Matsumoto-cho, Kasugai 487-8501, Japan, \email{ozawa@isc.chubu.ac.jp} \\
Graduate School of Informatics, Nagoya University,
Chikusa-ku, Nagoya 464-8601, Japan, \email{ozawa@is.nagoya-u.ac.jp}}
\maketitle
\abstract{ In this chapter, we present a general theory of finite quantum measurements, for which we assume that the state space of the measured system is a finite dimensional Hilbert space and that the possible outcomes of a measurement are a finite set of real numbers. We develop the theory in a deductive manner from the basic postulates for quantum mechanics and a few plausible axioms for general quantum measurements. We derive an axiomatic characterization of all the physically realizable finite quantum measurements. Mathematical tools necessary to describe measurement statistics such as POVMs and quantum instruments are not assumed at the outset, but we introduce them as natural consequences of our axioms. Our objective is to show that those mathematical tools can be naturally derived from obvious theoretical requirements.} 

\section{Introduction}

Theories in physics are closely related to mathematics.
In most cases, their basic principles are given undisputed 
mathematical expressions.
Hilbert's sixth problem explicitly asked to treat such mathematical
formulations of physical theories in a rigorous axiomatic method
that was successful in the investigations on the foundations of geometry
\cite{Hil03}.
One of the typical outcomes of this problem is von Neumann's axiomatization of 
quantum mechanics given in his 
book ``Mathematical Foundations of Quantum Mechanics''  (English translation) \cite{vN32E} originally published in 1932.

In contrast to the prevailing view that his work had completed 
the axiomatization of non-relativistic quantum mechanics without
any superselection rules, von Neumann's axioms need to be
completed by revising his treatments on quantum measurements.
It is true that until the 1970s there were no problems in non-relativistic
quantum mechanics that could not be solved because of the deficiency 
of von Neumann's axioms.
However, the technological development for precision measurements based
on laser optics from the 1960s revealed that von Neumann's axioms were not sufficient.

In quantum mechanics, the notion of measurement plays an indispensable role.
Nevertheless, ``theory of measurement'' was left incomplete in von Neumann's
axiomatization.
Quantum mechanics as axiomatized by von Neumann was only justified in 
experiments in which the object is prepared in the known state, undergoes time
evolution with the known Hamiltonian, and then is subject to a measurement
of one of its observables.  Von Neumann's quantum postulates successfully predict
the probability distribution of the outcome of a measurement. 
However, they are not sufficient to predict the joint probability
distribution of the outcomes of successive measurements in time.

Experimental technology was not precise enough to carry out successive 
measurements for comparing with the theory for long.
However, the emerging laser technology in the 1960s
enabled us to control quantum states and measuring interactions 
precisely enough to make repeated or successive measurements.

For  the statistics of successive measurements, we need the notion of state changes caused 
by measurements, often called quantum state reductions,
which has been considered one of the most difficult notions in quantum mechanics.
In order to deal with quantum systems to be measured sequentially, 
we need a general notion of quantum state reductions and mathematical methods
to calculate them.  For this purpose, we have to mathematically characterize 
all of the possible state changes caused by the most general type of measurements.

Von Neumann \cite{vN32E} introduced the repeatability hypothesis to determine the quantum
state reduction, which implies that the state of the measured system
changes to the eigenstate of the measured observable corresponding to the outcome
of the measurements.
This principle uniquely determines the state change for measurements of non-degenerate
discrete observables, of which all the eigenspaces are one dimensional. 
For degenerate observables the eigenstate is not uniquely
determined.   Subsequently, L\"{u}ders \cite{Lud51} proposed the
projection postulate, called the von Neumann-L\"{u}ders projection postulate,
to determine the unique eigenstate.

A mathematical problem was considered to extend this notion to measurements of
observables with continuous spectra (continuous observables).
Nakamura and Umegaki \cite{NU62} pointed out that the unique state change described 
by von Neumann's repeatability hypothesis for measurements of non-degenerate observables
 is an instance of Umegaki's non-commutative extension of the notion of
 conditional expectations  \cite{Ume54}, 
 which is a projection map of the full operator algebra onto the subalgebra generated by the measured 
 observable.  They conjectured that von Neumann's repeatability hypothesis can be
extended to continuous observables by Umegai's notion of non-commutative
conditional expectations.
 However, Arveson \cite{Arv67} proved that the corresponding conditional 
 expectation does not exist if the measured observable has a continuous spectral.
 
 Based on the above results, Davies and Lewis \cite{DL70} proposed to abandon
 the repeatability hypothesis as the first principle, and introduced a general 
 mathematical tool to describe a general measurement, called an instrument, 
 which generalizes the notion of Umegaki ``conditional expectations'' \cite{Ume54},  
 the notion of ``operations'' described for discrete observables by Schwinger \cite{Sch59,Sch60a,Sch60b}
 and introduced in the context of algebraic quantum field theory by 
 Haag and Kastler \cite{HK64}, the notion of ``effects'' introduced by
 Ludwig \cite{Lud67,Lud68}, and the notion of ``POVMs'' introduced by Helstrom \cite{Hel76}.  
 The Davies-Lewis (DL) instruments were considered as a general mathematical framework
 that may describe all the possible quantum measurements.  However, the
 consistency problem remained.
 It was not clear whether the class of DL instruments is too general or not general enough 
 to describe quantum measurements, or whether every DL instrument 
 is consistent with other postulates for quantum mechanics, 
 or whether every DL instrument has a quantum mechanical model for a measuring process.
 
In 1986,  Yuen \cite{Yue87} explicitly proposed the problem to find a mathematical 
characterization of all the physically realizable quantum measurements,
and he conjectured that the DL instruments are too general.
Shortly before this proposal, the problem had been solved by the present author 
\cite{84QC} in1984, showing that all the physically realizable
quantum measurements are faithfully characterized by completely
positive instruments, the notion which modifies the notion of 
DL instruments by requiring complete positivity.
This completed von Neumann's axiomatization of quantum mechanics
by adding the most general measurement axiom, or the most general description 
of quantum state reductions,  consistent with the other von Neumann axioms 
for quantum mechanics.

In this chapter, we present a general theory of finite quantum measurements,
as an introduction to a general theory of quantum measurements developed as outlined above, 
in a form accessible without mathematics for operators in infinite dimensional Hilbert spaces
and probability theory for continuous random variables.
Thus, we assume that the state space of the measured system is a finite dimensional Hilbert space
and that the possible outcomes of a measurement is a finite set of real numbers.
We develop the theory in a deductive manner from the basic postulates for quantum 
mechanics and a few plausible axioms for general quantum measurements,
and we derive an axiomatic characterization of all of the physically realizable
finite quantum measurements. 
Mathematical tools necessary to describe measurement statistics, such as POVMs 
and quantum instruments, are not assumed at the outset, but we introduce
them as natural consequences of our intuitive axioms.  Our objective is to show 
that those mathematical tools can be naturally derived from obvious theoretical requirements.    

As a chapter of a Festschrift celebrating Andrei Khrennikov and 
the quantum-like revolution,  the author hopes that this work would help the readers 
to extend the scope of quantum measurement theory based on the notion of quantum 
instruments beyond quantum physics.

\section{Quantum Mechanics}

In this chapter, we consider quantum systems described by finite dimensional Hilbert spaces.
Mathematically, a finite dimensional Hilbert space is defined as any finite dimensional
linear space $\cH$ over the complex number field with inner product $(\xi,\et)$ defined
for all $\xi,\et\in\cH$, which we assume linear in $\et$ and conjugate linear in $\xi$
 following the physics convention.
 
Let  $\cH$  be a finite dimensional Hilbert space.  A linear operator on $\cH$ is
a linear mapping defined everywhere on $\cH$ with values in $\cH$.
Denote by $\cL(\cH)$ the space of linear operators on $\cH$.
The adjoint of a linear operator $A$ on $\cH$ is a linear operator $A^{\da}$ uniquely
determined by the condition $(\xi,A^{\da}\et)=(A\xi,\et)$ for all $\xi,\et\in\cH$.
A linear operator $A$ is said to be self-adjoint if $A^{\da}=A$, or equivalently
$(\xi,A\xi)$ is a real number for all $\xi\in\cH$.
A linear operator $A$ is said to be positive, in symbols $A\ge 0$, if $(\xi,A\xi)\ge 0$ for all $\xi\in\cH$.
The trace of a linear operator $A$ is defined by  $\Tr[A]=\sum_{j}(\ph_j,A\ph_j)$ for
any orthonormal basis $\{\ph_j\}$.\footnote{
The trace is independent of the choice of the orthonormal basis $\{\ph_j\}$.
To see this let $\{\xi_k\}$ be another orthonormal basis.  Then we have
$\Tr[A]=\sum_{j}(\ph_j,A\ph_j)=\sum_{j,k}(\ph_j,\xi_k)(\xi_k,A\ph_j)=
\sum_{j,k}(A^{\da}\xi_k,\ph_j)(\ph_j,\xi_k)=\sum_{k}(A^{\da}\xi_k,\xi_k)
=\sum_{k}(\xi_k,A\xi_k)$.  Thus, the trace is independent of the choice of the
orthonormal basis.}
A linear operator $\rh$ is called a density operator if positive and of unit trace, i.e., 
$\rh\ge 0$ and $\Tr[\rh]=1$.  

Axioms for quantum mechanics of finite level systems without any superselection rules 
are given as follows.
\bigskip

{\bf Axiom Q1 (Quantum systems, states, and observables).}
{\em Every quantum system $\bS$ is described by a finite dimensional 
Hilbert space ${\mathcal H}$ called the {\em state space} of $\bS$. 
{\em States} of\/ $\bS$ are represented by density
operators on $\cH$ and {\em observables} 
of $\bS$ are represented by self-adjoint operators on $\cH$.
Every density operator on $\cH$ corresponds to a state of\/ $\bS$,
and every self-adjoint operator corresponds to an observable of\/ $\bS$.}
\bigskip

By Axiom Q1 we shall identify states with density operators,
and observables with self-adjoint operators.  The state of the form
$\rh=\ket{\ps}\bra{\ps}$ is called a {pure state}.\footnote{
For any $\xi,\et\in\cH$,  the operator $\ket{\xi}\bra{\et}$ is defined by
$(\ket{\xi}\bra{\et})\ps=(\et,\ps)\xi$ for all $\ps\in\cH$.
} 
If $\bS$ is in the state $\rh=\ket{\ps}\bra{\ps}$,
$\bS$ is said to be in the (vector) state $\ps$.
We denote by $\cS(\cH)$ the space of states, or density operators,  on $\cH$
and by $\cO(\cH)$ the space of observables, or self-adjoint operators,  on $\cH$. 
\bigskip

{\bf Axiom Q2 (Born statistical formula).}
{\em If an observable $A$ is measured in a state $\rho$, 
the outcome obeys the probability distribution
of $A$ in $\rho$ defined by the Born statistical formula (BSF) 
\beql{BSF}
\Pr\{\bx=x\|\rh\}=\Tr[P^{A}(x)\rh]
\eeq
where $x\in\R$, and $P^{A}(x)$ stands for the projection 
onto the subspace $\{\ps\in\cH| A\ps=x\ps\}$.}
\bigskip

The projection $P^{A}(x)$ is called the spectral projection of $A$
for the real number $x$.
The map $P^{A}:x\mapsto P^{A}(x)$ is called the 
spectral measure of $A$ \cite{Hal51}.  From Axiom Q2, the mean value  $\bracket{A}$
and the standard deviation $\si(A)$ are given by
\beqa
\bracket{A}&=&\Tr[A\rh],\\
\si(A)^2&=&\bracket{A^2}-\bracket{A}^2.
\eeqa
The standard deviations $\si(A),\si(B)$
of observables $A,B$ in a state $\rh$ satisfy 
Robertson's inequality
\beql{URSP}
\si(A)\si(B)\ge\frac{1}{2}|\bracket{[A,B]}|.
\eeq
\bigskip

{\bf Axiom Q3 (Time evolution).}
{\em Suppose that a system $\bS$ is an isolated system with the (time-independent)
Hamiltonian $H$ from time $t$ to $t+\ta$.
The system $\bS$ is in a state  $\rh(t)$ at time $t$ if and only if\/ $\bS$ is in the
state $\rh(t+\ta)$ at time $t+\ta$ satisfying
\beqa
\rh(t+\ta)=e^{-i\ta H/\hslash}\rh(t)e^{i\ta H/\hslash},
\eeqa
where $\hslash$ stands for the Planck constant divided by $2\pi$.}
\bigskip

{\bf Axiom Q4 (Composite systems).}
{\em The state space of the composite system $\bS=\bS_{1}+\bS_{2}$ 
of a system $\bS_{1}$ with the
state space $\cH$ and a system  $\bS_{2}$ with the
state space $\cK$ is given by the tensor product 
$\cH\otimes\cK$.  The observable $A$ of\/ $\bS_1$
is identified with the observable  $A\otimes I$  of\/ $\bS$ and 
the observable $B$ of\/ $\bS_2$ is identified with 
 $I\otimes B$ of \/ $\bS$.}
\bigskip

For any orthonormal bases $\{\xi_j\}$ of $\cH$ and $\{\et_k\}$ of $\cK$, 
the family of their tensor products
$\{\xi_j\otimes\et_k\}$ forms an orthonormal basis of $\cH\otimes\cK$.
The tensor product of $A\in\cL(\cH)$ and  $B\in\cL(\cK)$ is the linear operator 
$A\otimes B\in\cL(\cH\otimes\cK)$ defined by 
$(A\otimes B)(\xi\otimes \et)=A\xi\otimes B\et$ for all $\xi\in\cH$ and $\et\in\cK$.
Every $A\in \cL(\cH\otimes\cK)$ is of the form $A=\sum_{j,k}C_j\otimes D_k$
where $C_j\in\cL(\cH)$ and $D_k\in\cL(\cK)$.
The partial traces $\Tr_{\cK}[A]$ over $\cK$ and $\Tr_{\cH}[A]$ over $\cH$ 
of $A\in\cL(\cH\otimes \cK)$ 
are defined by $\Tr_{\cK}[A]=\sum_{j,k}C_j\otimes \Tr[D_k]$ and
$\Tr_{\cH}[A]=\sum_{j,k}\Tr[C_j]\otimes D_k$  if
$A=\sum_{j,k}C_j\otimes D_k$.
It follows from Axiom Q4 that 
if the system $\bS_{1}+\bS_{2}$ is in a state $\rh_{12}$, then the system
$\bS_{1}$ is in the state $\rh_{1}=\Tr_{\cK}[\rh_{12}]$  and $\bS_{2}$ is in the state 
$\rh_{2}=\Tr_{\cH}[\rh_{12}]$.

\section{Statistical properties of measuring apparatuses}

In this section, we discuss statistical properties of measuring apparatuses.
We introduce plausible axioms for statistical properties of a measuring apparatus
required for every apparatus to satisfy, and then we show that  statistical properties 
of physically realizable measuring apparatuses can be naturally described by POVMs
and completely positive instruments.

\subsection{Output probability distributions}

{\bf Axiom M1 (Output probability distributions and quantum state reductions).}
{\em An apparatus $\bA(\bx)$ with output variable $\bx$ to measure a system $\bS$ determines the probability 
$\Pr\{\bx=x\|\rh\}$ of  the outcome $\bx=x$ of the measurement depending 
on the input state $\rh$ (the state of\/ $\bS$ just before the measurement),
and determines the output state  $\rh_{\{\bx=x\}}$ (the state of\/ $\bS$ just after the measurement) 
depending on the input state $\rh$
and the outcome $\bx=x$ of the measurement.}
\bigskip

The variable representing the outcome of the apparatus is called the 
{\em output variable}.
Let ${\bf S}$ be a quantum system, to be referred to the {\em object}, 
described by a Hilbert space ${\cal H}$ of state vectors.
Let $\bA(\bx)$ be a {\em measuring apparatus} with an output
variable $\bx$ to measure the {object} $\bS$.
We assume that $\bx$ takes values in the real line $\R$.
For any real number  $x\in\R$, we shall denote by ``$\bx=x$'' 
the probabilistic event
that the output variable $\bx$ of apparatus $\bA(\bx)$ takes the value $x$.
By Axiom M1, the probability distribution of the output variable 
$\bx$ is determined by the input state $\rh$.
Denote it by $\Pr\{\bx=x\|\rh\}$, and call it as 
the {\em output distribution} of $\bA(\bx)$  in the input state $\rh$.
 If the state $\rh$ is a vector state $\rh=\ket{\ps}\bra{\ps}$,
we also write $\Pr\{\bx=x\|\ps\}=\Pr\{\bx=x\|\rh\}$.
For any subset $\De\subseteq \R$, we define 
$\Pr\{\bx\in\De\|\rh\}=\sum_{x\in\De}\Pr\{\bx=x\|\rh\}$.
We suppose that the output distribution satisfies the following conditions. 

(i) {\bf Positivity:} $\Pr\{\bx=x\|\rh\}\ge0$ for every $x\in\R$.

(ii) {\bf Unity}: $\sum_{x\in\R}\Pr\{\bx=x\|\rh\}=1$.

(iii) {\bf Finiteness}: There exists a finite subset of $S\subseteq\R$
such that if $x\not\in S$ then $\Pr\{\bx=x\|\rh\}=0$
for any state $\rh$.

The output probability distribution should satisfy the following
postulate.
\bigskip

{\bf Axiom M2' (Mixing law of output probability).}
{\em For any apparatus $\bA(\bx)$, the
function $\rh\mapsto\Pr\{\bx=x\|\rh\}$ is
an affine function of states $\rh$ for any real number $x$.}
\bigskip

This means that we have 
$$
\Pr\{\bx=x\|p\rh_{1}+(1-p)\rh_{2}\}
=p\Pr\{\bx=x\|\rh_{1}\}+(1-p)\Pr\{\bx=x\|\rh_{2}\},
$$
where $\rh_{1}$ and $\rh_{2}$ are density operators and $0<p<1$. 

The above axiom is justified by the following interpretation of
the mixture of states:  The system $\bS$ is in the state
$p\rh_{1}+(1-p)\rh_{2}$ if  it is in state $\rh_{1}$ with
probability $p$ and in state $\rh_{2}$ with probability $1-p$.

\subsection{Probability operator-valued measures}

In order to characterize output probability distributions, we introduce a mathematical
definition.  A mapping $\Pi:x \mapsto \Pi(x)$ 
of $\R$ into the space  $\cL(\cH)$ of linear operators on 
$\cH$ is called a
{\em probability operator-valued measure (POVM)}, also known as a {\em
positive operator-valued measure} \cite{NC00} or a {\em probability operator measure (POM)}
\cite{Hel76}, if the following conditions are satisfied: 

(i) {\bf Positivity}: $\Pi(x)\ge 0$ for all
$x\in\R$. 

(ii) {\bf Unity}: $\sum_{x\in\R}\Pi(x)=1$. 

(iii) {\bf Finiteness}: There exists a finite subset of $S\subseteq\R$
such that if $x\not\in S$ then $\Pi(x)=0$.

One of important consequences from the above postulate is
the following characterization of output probability 
distributions \cite{80OG}.

\begin{Theorem}\label{th:POVM}
Axiom M2' (the mixing law of output probability) holds if and only if
for any apparatus $\bA(\bx)$, there  uniquely exists  a 
POVM $\Pi$ satisfying
\beql{Born-POVM}
\Pr\{\bx=x\|\rh\}
=
\Tr[\Pi(x)\rh]
\eeq
for any real number $x$ and  density operator $\rh$.
\end{Theorem}

The POVM $\Pi$ defined by \Eq{Born-POVM} is called the {\em POVM of
$\bA(\bx)$}.  

\subsection{The Born statistical formula}

Let $A$ be an observable of system $\bS$.
According to Axiom Q2 (Born statistical formula),
we say that apparatus $\bA(\bx)$ {\em satisfies the Born statistical
formula (BSF)} for observable $A$ on input state $\rh$, if we have
\beq
\Pr\{\bx=x\|\rh\}=\Tr[P^{A}(x)\rh]
\eeq
for every real number $x$, 
where  $P^{A}(x)$ is the spectral projection of $A$ for $x$.  From Eqs.~\eq{Born-POVM} and \eq{BSF}, apparatus $\bA(\bx)$  
satisfies the Born statistical formula on every input state if and only if 
the POVM $\Pi$ of $\bA(\bx)$ is the spectral measure $P^{A}$.
In this case, the apparatus $\bA(\bx)$ is called an {\em $A$-measuring apparatus}.
Naturally, we assume that {\em for every observable $A$ of $\bS$ there 
is at least one $A$-measuring apparatus.}

The Born statistical formula is a necessary condition for the state-dependent
accuracy of measurement, or for the accurate measurements of an observable
$A$ in a single state $\rh$.
For a necessary and sufficient condition for the state-dependent accuracy of 
measurement, we refer the reader to \cite{05PCN,06QPC,19A1}.
According to those studies of the state-dependent accuracy of 
measurement, an apparatus accurately measures 
an observable $A$ in every state $\rh$ if and only if the Born statistical 
formula holds for every state $\rh$.  Thus, the state-dependent accuracy 
of measurement required for every state is consistent with the conventional 
state-independent accuracy of measurement.

\subsection{Quantum state reductions}
According to Axiom M1, 
depending on the input state $\rh$, 
any apparatus $\bA(\bx)$ determines the state 
$\rh_{\{\bx=x\}}$ just after the measurement for any possible 
outcome $\bx=x$ with $\Pr\{\bx=x\|\rh\}>0$. 
The operational meaning of the state $\rh_{\{\bx=x\}}$ is 
given as follows.
If a measurement using the apparatus $\bA(\bx)$ on input state $\rh$ 
is immediately followed by a measurement using another apparatus 
$\bA(\by)$ with output variable $\by$, we shall denote by
$\Pr\{\by=y|\bx=x\|\rh\}$ the conditional probability
of the outcome $\by=y$ of the measurement using $\bA(\by)$ given
the outcome $\bx=x$ of the measurement using $\bA(\bx)$.  
Then under the condition $\bx=x$
the state just before the measurement using $\bA(\by)$ is
the state $\rh_{\{\bx=x\}}$ so that we naturally have
\beql{020603a}
\Pr\{\by=y|\bx=x\|\rh\}=\Pr\{\by=y\|\rh_{\{\bx=x\}}\}.
\eeq
If $\Pr\{\bx=x\|\rh\}=0$, the state 
$\rh_{\{\bx=x\}}$ is taken to be indefinite.
The state $\rh_{\{\bx=x\}}$ is called the {\em output
state of the apparatus $\bA(\bx)$ given 
the outcome $\bx=x$ on input state $\rh$}.

Two apparatuses are called {\em
statistically equivalent} if they have the same
output probabilities and the same output states
for any outcomes and any input states.

\subsection{Joint output probability distributions}

If a measurement using apparatus $\bA(\bx)$ on input state $\rh$ 
is immediately followed by a measurement  using apparatus 
$\bA(\by)$, then from \Eq{020603a}
the joint probability distribution 
$\Pr\{\bx=x,\by=y\|\rh\}$ 
of the output variables $\bx$ and
$\by$ is given by
\beq\label{eq:c}
\Pr\{\bx=x,\by=y\|\rh\}
=\Pr\{\by=y\|\rh_{\{\bx=x\}}\}\Pr\{\bx=x\|\rh\}.
\eeq
Thus, the joint probability distribution of outputs of  successive
measurements depends only on the input state of the first
measurement. 
We shall call the above joint probability distribution the {\em joint output
probability distribution of $\bA(\bx)$ followed by $\bA(\by)$.}
The joint output probability distributions should satisfy the following
axiom:
\vskip\topsep

{\bf Axiom M2 (Mixing law of joint output probability). }
{\em For any apparatuses $\bA(\bx)$ and $\bA(\by)$, 
the function $\rh\mapsto \Pr\{\bx=x,\by=y\|\rh\}$
is an affine function of density operators $\rh$ for any real numbers $x,y$.}
\bigskip

Since the joint probability $\Pr\{\bx=x,\by=y\|\rh\}$ depends on the initial input
state $\rh$, the above axiom is also justified by the following interpretation of
the mixture of states:  The system $\bS$ is in the state
$p\rh_{1}+(1-p)\rh_{2}$ if  it is in state $\rh_{1}$ with
probability $p$ and in state $\rh_{2}$ with probability $1-p$.

By summing up $y$ in \Eq{c} and using the unity relation
$\sum_{y\in\R}\Pr\{\by=y\|\rh_{\{\bx=x\}}\}=1$, we have
\beq
\sum_{y\in\R}\Pr\{\bx=x,\by=y\|\rh\}=\Pr\{\bx=x\|\rh\}
\eeq
for any $x$ and $\rh$.  Thus, we conclude that {\em Axiom M2 (the mixing law
of joint output probability) implies Axiom M2' (the mixing law of 
output probability).}

\subsection{Instruments}

Davies and Lewis \cite{DL70,Dav76} 
introduced the following mathematical notion for unified 
description of statistical properties of measurements.
A mapping  $\cI:x\mapsto\cI(x)$  of $\R$ 
into the space $\cL(\tc(\cH))$ of linear transformations
on $\tc(\cH)$ is called a {\em Davies-Lewis (DL) instrument}, if the following conditions 
are satisfied.

 (i) {\bf Positivity:}
$\cI(x)$ maps positive operators in $\tc(\cH)$ to positive operators
in $\tc(\cH)$ for any$x\in\R$.

 (ii) {\bf Unity:}  $\sum_{x\in\R}\cI(x)$ is trace-preserving.

 (iii) {\bf Finiteness: } There exists a finite subset $S\subset \R$
 such that $\cI(x)=0$ for all $x\in S$.
 
For any apparatus $\bA(\bx)$, 
we define the mapping $\cI(x):\rh\mapsto \cI(x)\rh$
by 
\beql{instruments}
\cI(x)\rh=\Pr\{\bx=x\|\rh\}\rh_{\{\bx=x\}},
\eeq
where $\rh\in\cS(\cH)$ and $x\in\R$.
The mapping $\cI(x)$  transforms any density operator $\rh$ to
a positive operator with the trace
equal to $\Pr\{\bx=x\|\rh\}$.
It follows from Axiom M2 (the mixing law of joint output probability) that $\cI(x)$
is an affine mapping and can be extended to a linear 
transformation on the space $\tc(\cH)$ of linear operators 
on $\cH$ \cite{97OQ,00MN}.
Then it is easy to see that the mapping $\cI(x):\rh\mapsto \cI(x)\rh$ satisfies the
Davies and Lewis definition of instruments.
Conversely, if any apparatus $\bA(\bx)$ has a Davies-Lewis instrument $\cI$ 
satisfying \Eq{instruments}, then Axiom M2 (the mixing law of joint output probability)
holds. 
Thus, we have \cite{00MN} 

\begin{Theorem}
\label{th:instrument}
Axiom M2 (the mixing law of joint output probability) holds if and only if 
for any apparatus $\bA(\bx)$ there uniquely exists 
a DL instrument $\cI$ satisfying \Eq{instruments}
for any real number $x$ and  density operator $\rh$.
\end{Theorem}

The mapping  $\cI(x)$ defined by \Eq{instruments} for  the apparatus $\bA(\bx)$ 
is called 
the {\em operation of $\bA(\bx)$ given the outcome  $\bx=x$}. The mapping $\cI$
is called the {\em instrument of $\bA(\bx)$}.
Then the output probability and the output state can be
expressed by
\beqa
\Pr\{\bx=x\|\rh\}&=&\Tr[\cI(x)\rh],
\label{eq:DL1}\\
\rh_{\{\bx=x\}}&=&\frac{\cI(x)\rh}{\Tr[\cI(x)\rh]}
\label{eq:DL2}
\eeqa
where the second relation assumes $\Pr\{\bx=x\|\rh\}>0$.
Thus, if $\cI_{\bx}$ and $\cI_{\by}$ are the instruments
of $\bA(\bx)$ and $\bA(\by)$, respectively, 
then the joint output probability distribution can be expressed by
\beql{JOPD1}
\Pr\{\bx=x,\by=y\|\rh\}
=\Tr[\cI_{\by}(y)\cI_{\bx}(x)\rh]
\eeq
for any state $\rh$ and any real numbers $x,y$.

Both the output probability distribution and the output states
are determined by the instrument.  Thus,
two apparatuses are statistically equivalent if and only if 
they have the same instrument.

For any linear transformation $T$ on the
space $\tc(\cH)$ of linear operators on $\cH$,
the {\em dual} of $T$ is defined to be the linear
transformation $T^{*}$ on  $\cL(\cH)$ satisfying 
\beql{dual}
\Tr[A(T\rh)]=\Tr[(T^{*}A)\rh]
\eeq
for any $A, \rh\in\cL(\cH)$. 
The dual  of the operation $\cI(x)$ is called the 
{\em dual operation $\cI(x)^{*}$ given  $\bx=x$};
by \Eq{dual} it is defined by the relation
\beql{dual-operation}
\Tr[A\cI(x)\rh)]=\Tr\{[\cI(x)^{*}A]\rh\}
\eeq
for any $A, \rh\in\cL(\cH)$.

The operator $\cI(x)^{*}I$ obtained by applying
the dual operation $\cI(x)^{*}$ to the identity operator $I$
is called the {\em effect of the operation $\cI(x)$}.
By \Eq{DL1} and \Eq{dual} we have
\beq
\Pr\{\bx=x\|\rh\}=\Tr[(\cI(x)^{*}I)\rh].
\eeq
Since $\rh$ is arbitrary, comparing with \Eq{Born-POVM},
we have
\beql{Inst-POVM}
\Pi(x)=\cI(x)^{*}I
\eeq
for any real number $x$.  Thus, the POVM of $\bA(\bx)$ is
determined by the effects of the instrument $\cI$.

Let $\cI_{\bx}$ and $\cI_{\by}$ be the instruments of
$\bA(\bx)$ and $\bA(\by)$, respectively, and let $\Pi_{\by}$ be the
POVM of $\bA(\by)$. 
Then we have

\beqa
\Tr[\cI_{\by}(y)\cI_{\bx}(x)\rh]&=&
\Tr\{[\cI_{\by}(y)^{*}I][\cI_{\bx}(x)\rh]\}\nn\\
&=&
\Tr\{[\Pi_{\by}(y)[\cI_{\bx}(x)\rh]\}\nn\\
&=&
\Tr\{[\cI(x)^{*}\Pi_{\by}(y)]\rh\}
\eeqa
Thus, the joint output probability distribution can be expressed by
\beql{JOPD2}
\Pr\{\bx=x,\by=y\|\rh\}
=\Tr\{[\cI(x)^{*}\Pi_{\by}(y)]\rh\}
\eeq
for any $x,y\in\R$.

\subsection{Selective quantum state reduction}

For any subset $\De$ of $\R$, the outcome event ``$\bx\in\De$'' 
means that the outcome of the measurement is an element of 
$\De$.  The probability of outcome event $\bx\in\De$ is given by
\beq
\Pr\{\bx\in\De\|\rh\}=\sum_{x\in\De}\Pr\{\bx=x\|\rh\}.
\eeq
If the input state is  $\rh$, the state just after the measurement given
 the outcome event $\bx\in\De$ 
is denoted by $\rh_{\{\bx\in\De\}}$.
This state is determined as follows.
Let $A$ be an observable of the object $\bS$.
Suppose that the observer measures the object $\bS$ in the state $\rh_{\{\bx\in\De\}}$ 
using  another apparatus   $\bA(\by)$  with the POVM $\Pi_{\by}=P^{A}$. 
Then we have
\beqa
\Pr\{\bx\in\De,\by=y\|\rh\}&=&\Pr\{\by=y\|\rh_{\{\bx\in\De\}}\}\Pr\{\bx\in\De\|\rh\}\nn\\
&=&\Tr[P^{A}(y)\Pr\{\bx\in\De\|\rh\}\rh_{\{\bx\in\De\}}]\label{eq:110602a}
\eeqa
On the other hand,  by \Eq{c} we have
\beqa
\Pr\{\bx\in\De,\by=y\|\rh\}
&=&\sum_{x\in\De}\Pr\{\bx=x,\by=y\|\rh\}\nn\\
&=&\sum_{x\in\De}\Pr\{\by=y\|\rh_{\{\bx=x\}}\}\Pr\{\bx=x\|\rh\}\nn\\
&=&\Tr[P^{A}(y)\sum_{x\in\De}\Pr\{\bx=x\|\rh\}\rh_{\{\bx=x\}}].
\label{eq:110602b}
\eeqa
Since $A$ is an arbitrary observable, 
 by comparing \Eq{110602a} and \Eq{110602b},
we have 
\beq
\Pr\{\bx\in\De\|\rh\}\rh_{\{\bx\in\De\}}=
\sum_{x\in\De}\Pr\{\bx=x\|\rh\}\rh_{\{\bx=x\}}.
\eeq

For any subset $\De$ of $\R$, we write
 $\cI(\De)=\sum_{x\in\De}\cI(x)$ and $\Pi(\De)=\sum_{x\in\De}\Pi(x)$.
Let  $\cI$  be the instrument of an apparatus  $\bA(\bx)$.
For any state  $\rh$ we have
\beq
\cI(\De)\rh=\Pr\{\bx\in\De\|\rh\}\rh_{\{\bx\in\De\}}.
\eeq
$\cI(\De)$ is called {\em the operation given the outcome event
$\bx\in\De$} of the apparatus $\bA(\bx)$.

The state change from the state  $\rh$ to the state $\rh_{\{\bx=x\}}$ 
is called an {\em (individual) quantum state reduction}.
The state change from the state  $\rh$ to the state $\rh_{\{\bx\in\De\}}$ 
is called a {\em selective quantum state reduction}.
On the other hand, the state change $\rh\mapsto\rh_{\{\bx\in\R\}}$ is called 
a {\em  non-selective quantum state reduction}.
For the instrument $\cI$  of the apparatus $\bA(\bx)$,
the operation  $T=\cI(\R)$ is called the {\em non-selective operation}  of $\bA(\bx)$,
and $T^{*}=\cI(\R)^{*}$ is called  the  {\em  non-selective dual operation} of 
$\bA(\bx)$.
In general a linear transformation $T$ on $\cL(\cH)$ is called a {\em positive map}
if $T\rh\ge 0$ for all $\rh\ge 0$.
For any DL instrument $\cI$, $\cI(\De)$ is a positive map.
The non-selective operation $T$  is trace-preserving, i.e., 
\beq
\Tr[T\rh]=\Tr[\rh],
\eeq
for any $\rh\in\cL(\cH)$,
while the non-selective dual
operation $T^{*}$ is unit-preserving, 
\beq
T^{*}I=I.
\eeq

\subsection{Repeatability Hypothesis}

In the early days of quantum mechanics,
only a restricted class of measurements was seriously studied.
The following axiom was broadly accepted in the 1930s.
\bigskip

{\bf (M) Measurement axiom.}
If an observable $A$ is measured in a system $\bS$ to obtain the outcome $a$, 
then the system $\bS$ is left in an eigenstate of $A$ for the eigenvalue $a$.
\bigskip

Von Neumann \cite{vN32E} showed that this assumption is equivalent to the following
assumption called the {\em repeatability hypothesis} \cite[p.~335]{vN32E}, 
posed with a clear operational condition generalizing a feature of the Compton-Simons 
experiment \cite[pp.~212--214]{vN32E}. 
\bigskip

{\bf (R) Repeatability hypothesis.}
{\em If an observable $A$ is measured twice in succession 
in a system $\bS$, then we get the same value each time.}
\bigskip

It can be seen from the following definition of measurement due to Schr\"{o}dinger
given in his famous ``cat paradox'' paper \cite{Sch35} that 
von Neumann's repeatability hypothesis was broadly accepted in the 1930s.

\begin{quote}
The systematically arranged interaction of two systems (measured object and 
measuring instrument) is called a measurement on the first system, if a directly-sensible variable
feature of the second (pointer position) is always reproduced within certain error limits when the
process is immediately repeated (on the same object, which in the meantime must not be exposed 
to any additional influences) \cite{Sch35}.
\end{quote}

The repeatability hypothesis uniquely determines the state 
after the measurement if the measured observable $A$ is non-degenerate, i.e., 
every eigensubspace is one-dimensional.
Let $A=\sum_{n}a_n\ketbra{\ph_n}$ be a non-degenerate observable,
where $\ketbra{\ph_n}$ stands for the projection onto the subspace spanned by
the eigenvector $\ph_n$ for the eigenvalue $a_n$.
Then the measuring apparatus satisfies the repeatability hypothesis
if and only if the corresponding instrument is of the form:
\beq
\cI(x)\rh=\ketbra{\ph_n}\rh\ketbra{\ph_n}
\eeq
for any $\rh\in\cS(\cH)$ if $x=a_n$; and $\cI(x)=0$ otherwise.

If the measured observable is degenerate, the repeatability hypothesis 
does not determine the unique eigenstate as the state after the measurement.
L\"{u}ders \cite{Lud51} proposed the projection postulate to determine
the eigenstate uniquely.
\bigskip

 {\bf  (P) The von Neumann-L\"{u}ders projection postulate}.
 If a measurement of an observable $A$ in a state 
$\rh$ leads to the outcome $\bx=x$, the state  $\rh_{\{\bx=x\}}$  just after the measurement is given by
\beqa
\rh_{\{\bx=x\}}=\frac{P^{A}(x)\rh P^{A}(x)}{\Tr[P^{A}(x)\rh]}.
\eeqa
\bigskip

Thus, the projection postulate uniquely determines the instrument for
measurement of $A$ as
\beq
\cI(x)\rh=P^{A}(x)\rh P^{A}(x)
\eeq
for all $x\in\R$ and $\rh\in\cS(\cH)$.

It is well known that the same observable can be measured with many different ways
that do not satisfy the projection postulate.  Thus, the von Neumann-L\"{u}ders projection 
postulate should not be taken as a universal postulate for quantum mechanics 
but should be taken as a defining condition for a class of measurements called 
{\em projective measurements}.

For any sequence of projective measurements, we can determine the joint probability distribution 
of the outcomes of measurements \cite{Wig63+}.

\begin{Theorem}[Wigner formula]
\sloppy
Let $A_{1},\ldots,A_{n}$ be observables of a system $\bS$ in a state $\rh$ at time 0.
If one carries out projective measurements 
of observables $A_{1},\ldots,A_{n}$ at times 
$(0<)t_{1}<\cdots<t_{n}$ and otherwise leaves the system $\bS$ isolated
with the Hamiltonian $H$,
then the joint probability distribution of the outcomes 
$\bx_{1},\ldots,\bx_{n}$ of those measurements is given by
\beqa
\Pr\{\bx_{1}=x_{1},\ldots,\bx_{n}= x_{n}\|\rh\}
&=&
\Tr[P^{A_{n}}(x_{n})\cdots U(t_{2}-t_{1})P^{A_{1}}(x_{1})U(t_{1})\rh\nn
\\
& &\mbox{ }\times
U(t_{1})^{\da}P^{A_{1}}(x_{1})U(t_{2}-t_{1})^{\da}
\cdots P^{A_{n}}(x_{n})],\nn\\
\eeqa
where $U(t)=e^{-iHt/\hslash}$.
\end{Theorem}

\subsection{Abandoning the Repeatability Hypothesis}
\label{se:ABRH}

The repeatability hypothesis applies only to a restricted class of measurements 
and does not generally characterize the state changes caused 
by quantum measurements.
In fact, there exist commonly used measurements of discrete observables,
such as photon counting, that do not satisfy the repeatability hypothesis \cite{IUO90}.
Moreover,  it has been shown that the repeatability hypothesis cannot be generalized 
to continuous observables in the standard formulation of quantum mechanics
\cite{84QC,85CA,Sri80,88MR}.
In 1970, Davies and Lewis \cite{DL70} proposed abandoning the repeatability
hypothesis and introduced a new mathematical framework to treat 
all the physically realizable quantum measurements.
\begin{quote}
One of the crucial notions is that of repeatability which we show is implicitly
assumed in most of the axiomatic treatments of quantum mechanics, but whose
abandonment leads to a much more flexible approach to measurement theory
\cite[p.~239]{DL70}.
\end{quote}

The proposal of Davies and Lewis \cite{DL70} can be stated as follows.
\bigskip

{\bf  (DL) The Davies-Lewis thesis.}
{\em For every measuring apparatus $\bA(\bx)$ with output variable $\bx$ 
there exists  a unique DL instrument
$\cI$ satisfying
\beqa
\Pr\{\bx=x\|\rh\}&=&\Tr[\cI(x)\rh],\\
\rh\to\rh_{\{\bx=x\}}&=&
\frac{\cI(x)\rh}{\Tr[\cI(x)\rh]}.
\eeqa}
\bigskip

We have previously shown that under Axiom M1, the Davies-Lewis thesis is 
equivalent to Axiom M2 (the mixing law of joint output probability).

\subsection{Complete positivity}

Is every Davies-Lewis instrument physically relevant?
We shall show that this is not the case.
In physics various phenomena can be described by mathematical models,
and even in a single physical theory a single phenomenon can be modeled
by various different mathematical models.   
Nevertheless, we should have consistency relations among all the models
describing a single physical phenomenon such as invariance under the 
change of coordinate systems.
In quantum measurement theory, a single measuring apparatus can have
different models even with a fixed coordinate system, according to the
arbitrariness of the spacial boundary of the measured object.
As described by Axiom Q4, the observable $A$ in a system $\bS$ can 
have a different mathematical representative $A\otimes I$ in a larger system
$\bS+\bS'$.
Thus, an apparatus measuring the observable $A$ is accompanied by another model
describing it as an apparatus measuring the observable $A\otimes I$.
Namely, any model of an apparatus measuring the system $\bS$ is always  accompanied
by the model for an apparatus measuring the system $\bS+\bS'$.
It is interesting that this rather obvious fact leads to an important common 
property of measuring apparatuses.
\bigskip

{\bf Axiom M3 (Extendability).}
{\em For any apparatus $\bA(\bx)$ measuring a system $\bS$
and any quantum system $\bS'$ not interacting with $\bA(\bx)$ 
nor $\bS$,  there exists an apparatus $\bA(\bx')$ measuring system
$\bS+\bS'$  with the following statistical properties:
\beqa\label{eq:ex-1}
\Pr\{\bx'=x\|\rh\otimes\rh'\}
&=&
\Pr\{\bx=x\|\rh\},\\
(\rh\otimes\rh')_{\{\bx'=x\}}&=&\rh_{\{\bx=x\}}\otimes
\rh'
\label{eq:ex-2}
\eeqa
for any $x,x'\in \R$, any state $\rh$ of $\bS$, 
and any state $\rh'$ of $\bS'$.}
\bigskip

The above postulate is justified as follows.
Suppose that apparatus $\bA(\bx)$ measures object $\bS$.
Let $\bS'$ be any system, described by a Hilbert space $\cH'$, 
remote from $\bS$ and
$\bA(\bx)$.
Naturally, the apparatus $\bA(\bx)$ makes no measurement on $\bS'$
but then the same apparatus can be formally described 
as an apparatus $\bA(\bx')$ measuring 
the system $\bS+\bS'$ with the following statistical properties:
\beqa
\Pr\{\bx'=x\|\rh\otimes\rh'\}
&=&
\Pr\{\bx=x\|\rh\},\\
(\rh\otimes\rh')_{\{\bx'=x\}}&=&\rh_{\{\bx=x\}}\otimes
\rh'
\eeqa
for any real number $x$, any state $\rh$ of $\bS$ and any state $\rh'$ of
$\bS'$. 
Let $\cI'$ be the instrument of the apparatus $\bA(\bx')$.
Then we have
\beqa
\cI'(x)(\rh\otimes\rh')
&=&\Pr\{\bx'=x\|\rh\otimes\rh'\}(\rh\otimes\rh')_{\{\bx'=x\}}\nn\\
&=&\Pr\{\bx=x\|\rh\}\rh_{\{\bx=x\}}\otimes\rh'\nn\\
&=&[\cI(x)\rh]\otimes \rh'.
\eeqa
It follows that the operation $\cI'(x)$ of the extended apparatus
$\bA(\bx')$ given  $\bx'=x$ is represented by  
$\cI'(x)=\cI(x)\otimes \id $, where $\id$ stands for the identity map on $\cL(\cH')$.

Let $\cH'$ be a finite dimensional Hilbert space.
Any linear transformation $T$ on $\tc(\cH)$
can be extended naturally to the linear transformation $T\otimes {\rm id}_{\cH'}$
on  $\tc(\cH\otimes\cH')=\tc(\cH)\otimes\cL(\cH')$
by
\beq
(T\otimes \id)(\sum_{j}\rh_{j}\otimes \rh'_{j})
=\sum_{j}T(\rh_{j})\otimes\rh'_{j}
\eeq
for any $\rh_{j}\in\cL(\cH)$ and $\rh'_{j}\in\tc(\cH')$.
Then $T$ is called 
{\em completely positive (CP)}, if $T\otimes \id$ maps
positive operators in $\cL(\cH\otimes\cH')$ to positive
operators in  $\cL(\cH\otimes\cH')$ for any $\cH'$.
A DL instrument $\cI$ is called a {\em completely positive (CP) instrument},
if the operation $\cI(x)$ is CP for every $x\in\R$.

 Then from the positivity of the operation
$\cI'(x)$, the complete positivity of the original operation $\cI(x)$
follows.  Thus, Axiom M3 (Extendability) is equivalent to the following 
Axiom M3' (Complete positivity).
\bigskip

{\bf Axiom M3' (Complete positivity).}
{\em The instrument of every apparatus should be a CP instrument.}
\bigskip

We have posed two plausible requirements for the measurement statistics 
to be satisfied by any apparatus,
the mixing law of joint output probability distributions $\Pr\{\bx=x,\by=y\| \rh\}$ 
and the extendability of measurement statistics, 
as a set of necessary conditions for every apparatus to satisfy. 
Under these conditions, we have shown that every
apparatus corresponds uniquely to a CP instrument
that determines the output probability distributions and
the quantum state reduction caused by the apparatus.
Thus, the problem of determining physically  possible output probability distributions
and quantum state reductions is reduced to the problem as to which
CP instrument corresponds to a physically realizable apparatus.
This problem will be discussed in the next section and it will be shown 
that every CP instruments corresponds to a physically realizable apparatus. 

Now we note that there indeed exists a DL instrument that is not a CP instrument. 
The transpose operation of matrices in a fix basis is 
a typical example of a positive linear map which is not CP \cite{NC00}.
Let $T$ be a transpose operation on $\cL(\cH)$ 
for $\cH$,
and let $\mu(x)$ be any probability distribution supported in a finite subset of $\R$,
i.e., there exists a finite subset $S\subseteq\R$ such that $\mu(x)=0$ if $x\not\in S$.
Then the relation
\beqa
\cI(x)\rh=\mu(x)T(\rh)
\eeqa
for any $x\in\R$ and any operator $\rh$
defines a DL instrument.  
However, since $T$ is not CP, the operation
$\cI(x)$ is not CP, so the $\cI$ is not a CP instrument.
The extendability postulate implies that there is no physically realizable 
apparatus corresponding to the above instrument.

\section{Measuring processes}
In this section, we discuss measuring processes.
We introduce indirect measurement models, a class of universal models for 
measuring processes, carried out by physically realizable measuring apparatuses.
We analyze them according to quantum mechanics, and show that every 
indirect measurement models uniquely determines the instrument of the apparatus
that is a completely positive instrument, and conversely that every completely 
positive instrument is described by an indirect measurement model of an apparatus. 
Since any apparatus described by an indirect measurement model is considered
physically realizable, in principle, we conclude that an apparatus is physically realizable
if and only if its statistical properties are described by a CP instrument.

\subsection{Indirect measurement models}

Let $\bA(\bx)$ be a measuring apparatus with the macroscopic 
output variable $\bx$ to measure the object $\bS$.  
The measuring
interaction turns on at time $t$, the {\em time of measurement}, and turns off at
time $t+\De t$ between  object $\bS$ and apparatus $\bA(\bx)$.  
We assume that {\em the object and the apparatus do not 
interact each other before $t$ nor after $t+\De t$ and that 
the composite system $\bS+\bA(\bx)$ is isolated in the
time interval $(t,t+\De t)$}.  The {\em probe} $\bP$ is
defined to be the minimal part of apparatus $\bA(\bx)$ 
such that the composite system $\bS+\bP$ is isolated in the time interval
$(t,t+\De t)$. By minimality, we naturally assume that probe $\bP$ is a 
quantum system represented by a Hilbert space $\cK$.  
Denote by $U$ the unitary operator on $\cH\otimes\cK$ representing 
the time evolution of $\bS+\bP$ for the time interval $(t,t+\Delta t)$.  

At the time of measurement the object is supposed to
be in an arbitrary input state $\rh$
and the probe is
supposed to be prepared in a fixed state $\si$.
Thus, the composite system $\bS+\bP$ is in the state $\rh\otimes\si$
at time $t$ and in the state $U(\rh\otimes\si)U^{\da}$ at time $t+\De t$. 
Just after the measuring interaction, the object is separated from the apparatus,
and the probe is subjected to a local interaction with the subsequent stages of the
apparatus.  The last process is assumed to measure an observable $M$, called the
{\em meter observable}, of the probe, without further interacting with the object $\bS$,
and the output is represented by the value of the output variable $\bx$.  

\sloppy
Any physically realizable apparatus $\bA(\bx)$
can be modeled as above by a quadruple
$({\cal K},\si,U,M)$, called an {\em indirect measurement model}, 
of a Hilbert space
${\cal K}$,  a density operator $\si$ in ${\cal K}$, 
a unitary operator $U$ on ${\cal K}\otimes{\cal H}$,
and a self-adjoint operator $M$ on ${\cal K}$, where
${\cal K}$ represents the state space of the probe,
$\si$ the preparation of the probe,
$U$ the interaction between the object and the probe,
and $M$ the meter observable  to be detected.
An indirect measurement model $({\cal K},\si,U,M)$
is called {\em pure}, if $\si$ is a pure state; we shall
write $({\cal K},\si,U,M)=({\cal K},\xi,U,M)$, if
$\si=\ket{\xi}\bra{\xi}$.
 
\subsection{Output probability distributions}

Let $\bA(\bx)$ be an apparatus with
indirect measurement model
$(\cK,\si,U,M)$.
Since the outcome
of this measurement is obtained by the measurement of 
the meter observable  $M$ at time $t+\De t$, 
by the BSF \eq{BSF} for observable $M$ on input state $U(\rh\otimes\si)U^{\da}$
the output probability distribution of $\bA(\bx)$ is determined 
by
\beql{828e}
\Pr\{\bx=x\|\rh\}
=\Tr\{[I\otimes P^{M}(x)]U(\rh\otimes\si)U^{\da}\}.
	\eeq

By linearity of operators and the trace, it is easy to check that
the output probability distribution of $\bA(\bx)$ satisfies the
mixing law of output probability.
Thus, by Theorem \ref{th:POVM} there exists the POVM  $\Pi$ of 
$\bA(\bx)$.
To determine $\Pi$, 
using the partial trace operation $\Tr_{\cK}$ over $\cK$ 
we rewrite \Eq{828e} as
\beql{020604a}
\Pr\{\bx=x\|\rh\}
=\Tr[\Tr_{\cK}\{U^{\da}[I\otimes P^{M}(x)]U(I\otimes\si)\}\rh].
\eeq
Since $\rh$ is arbitrary, comparing Eqs.~\eq{Born-POVM} and
\eq{020604a}, POVM of $\bA(\bx)$ is determined  as
\beq
\Pi(x)=\Tr_{\cK}\{U^{\da}[I\otimes P^{M}(x)]U(I\otimes\si)\}
\eeq
for any $x\in\R$.

\subsection{Quantum state reductions}

Since the composite system $\bS+\bP$ is in the state
$U(\rh\otimes\si)U^{\da}$ at time $t+\De t$, from Axiom Q4
it is standard  that the object state at the time $t+\De t$
is obtained by tracing out the probe part of that state.
Thus, the nonselective state change is determined by
\begin{equation}  \rh\mapsto\rh'=
\Tr_{{\cal K}}[U(\rh\otimes\sigma)U^{\dagger}].
\end{equation}

In order to determine the quantum state reduction
caused by apparatus $\bA(\bx)$, 
suppose that at time $t+\Delta t$
the observer were to locally measure
an arbitrary observable $B$
of the same object ${\bf S}$.  
Let $\bA(\by)$ be a $B$-measuring apparatus
with output variable $\by$.
Since both the $M$ measurement on $\bP$ and the $B$ measurement
on $\bS$ at time $t+\Delta t$ are local, 
the joint probability distribution of their outcomes  
satisfies the joint probability formula for the
simultaneous measurement of $I\otimes M$ and 
$B\otimes I$ in the state 
$U(\rh\otimes\sigma)U^{\dagger}$ \cite{01OD}.

It follows that the joint output probability distribution of 
 $\bA(\bx)$ and $\bA(\by)$ is given by
\beq
{\rm Pr}\{\bx=x,\by=y\|\rh\}
=\Tr\{[P^{B}(y)\otimes P^{M}(x)]
U(\rh\otimes\sigma)U^{\dagger}\}.
\label{eq:215d}
\eeq
Thus, using the partial trace $\Tr_{\cK}$ we have
\beq
\Pr\{\bx=x,\by=y\|\rh\}
=\Tr[P^{B}(y) \Tr_{\cK}\{[I\otimes P^{M}(x)]U(\rh\otimes\sigma)U^{\dagger}\}].
\label{eq:020604b}
\eeq
On the other hand, from \Eq{c} 
the same joint output probability distribution can be represented by 
\begin{eqnarray}
{\rm Pr}\{\bx=x,\by=y\|\rh\}
&=&
{\Tr}[P^{B}(y)\rh_{\{\bx=x\}}]
{\Pr}\{\bx=x\|\rh\}\nn\\
&=&
{\Tr}[P^{B}(y){\Pr}\{\bx=x\|\rh\}\rh_{\{\bx=x\}}].
\label{eq:215e}
\end{eqnarray}

Since $B$ is chosen arbitrarily,
comparing Eqs.~\eq{020604b} and \eq{215e}, we have
\beql{020604c}
{\Pr}\{\bx=x\|\rh\}\rh_{\{\bx=x\}}=
\Tr_{\cK}\{[I\otimes P^{M}(x)]
U(\rh\otimes\sigma)U^{\dagger}\}.
\eeq    From \Eq{020604c}, the state 
$\rh_{\{\bx=x\}}$ is uniquely determined as
\begin{eqnarray}
\rh_{\{\bx=x\}}
&=&
\frac{\Tr_{{\cal K}}\{[I\otimes P^{M}(x)]
U(\rh\otimes\sigma)U^{\dagger}\}}
     {\Tr\{[I\otimes
P^{M}(x)]U(\rh\otimes\sigma)U^{\dagger}\}}.
\label{eq:215f}
\end{eqnarray}

By \Eq{020604c} the instrument $\cI$ of the apparatus $\bA(\bx)$
is determined  by
\beql{020609a}
\cI(x)\rh=\Tr_{{\cal K}}\{[I\otimes P^{M}(x)]
U(\rh\otimes\sigma)U^{\dagger}\}
\eeq
for any $x\in\R$ and any state $\rh$.  From the above relation, it is easy to see that
$\cI(x)$ satisfies the complete positivity;
as an alternative characterization, 
it is well-known that a linear transformation $T$ on
$\tc(\cH)$ is completely positive if and only if 
\beq
\sum_{ij}
(\xi_{i},T(\rh_{i}^{\da}\rh_{j})\xi_{j})\ge 0
\eeq
for any finite sequences $\xi_{1},\ldots,\xi_{n}\in\cH$ and 
$\rh_{1},\ldots,\rh_{n}\in\tc(\cH)$ \cite{84QC}.
In fact, we have
\beqas
\sum_{ij}
(\xi_{i},\cI(x)(\rh_{i}^{\da}\rh_{j})\xi_{j})
&=&\sum_{ij}\Tr[\cI(x)(\rh_{i}^{\da}\rh_{j})\ket{\xi_{j}}\bra{\xi_i}]\\
&=&\sum_{ij}\Tr\{[\ket{\xi_{j}}\bra{\xi_i}\otimes P^{M}(x)]U(\rh_{i}^{\da}\rh_{j}\otimes \si)U^{\da}\}\\
&=&\Tr[X^{\da}X]\ge 0,
\eeqas
where 
\beqas
X=\sum_{j}U(\rh_j\otimes \sqrt{\si})U^{\da}\ket{\xi_j}\bra{\ph}\otimes P^{M}(x)
\eeqas
for an arbitrary unit vector $\ph$.

Thus, we conclude that 
{\em the instrument of any apparatus
with indirect measurement  model
$({\cal K},\sigma,U,M)$ is a CP instrument}.  

The converse of this assertion was proven in \cite{84QC},
and hence we conclude this section by the following theorem.

\begin{Theorem}[Realization theorem]
\label{th:realization}
The instrument of any apparatus
with an indirect measurement model
is a CP instrument,
and conversely every CP instrument is obtained in this
way with a pure indirect measurement model.
\end{Theorem}

According to the above theorem, we conclude that an apparatus satisfying Axiom M1
is physically realizable if and only if it satisfied Axiom M2 (the mixing law of joint output 
probability) and Axiom M3 (the extendability). 

\section{General measurement axiom}
 For quantum systems with finite dimensional state spaces,
we can now complete von Neumann's axiomatization of quantum 
mechanics by augmenting it by the general measurement axiom that describes 
all the physically realizable measurements consistent with the other axioms
of quantum mechanics.
\bigskip

{\bf Axiom Q5 (General measurement axiom).}
{\em Every physically realizable apparatus $\bA(\bx)$ for the system $\bS$ with the
state space $\cH$ uniquely corresponds to a completely positive instrument $\cI$ for $\cH$
such that the statistical properties of $\bA(\bx)$ are determined by 
\beq
\cI(x)\rh=\Pr\{\bx=x\|\rh\}\rh_{\{\bx=x\}},
\eeq
or equivalently
\beqa
\Pr\{\bx=x\|\rh\}&=&\Tr[\cI(x)\rh], \\
\rh_{\{\bx=x\}}&=&\frac{\cI(x)\rh}{\Tr[\cI(x)\rh]},\quad\mb{if\/ $\Tr[\cI(x)\rh]>0$}
\eeqa
for all $x\in\R$ and $\rh\in\cS(\cH)$.  Conversely, every completely positive instrument $\cI$ for $\cH$
has at least one physically realizable apparatus $\bA(\bx)$ with the above
statistical properties.}
\bigskip

 According to the Kraus theorem \cite{Kra71}, for every completely positive map $T$ there exists a family of 
 operators $\{M_j\}$ on $\cH$ satisfying
 \beql{MO-1}
 T\rho=\sum_{j}M_j \rho M_j^{\da}
 \eeq
 for every $\rho\in\cL(\cH)$,
and 
\beql{MO-2}
T^{*}I=\sum_{j}M_j^{\da}M_j.
\eeq
It can be easily seen that every family of operators $\{M_j\}$ defines a completely positive map $T$ satisfying
Eqs.~\eq{MO-1}, \eq{MO-2}. 
It follows that for every completely positive instrument $\cI$ for $\cH$ there exits a family of operators $\{M_{xj}\}$
on $\cH$, called the {\em measurement operators} for $\cI$, satisfying
\beql{MO-4}
\cI(x)\rho=\sum_{j}M_{xj} \rho M_{xj}^{\da}
\eeq
 for every $\rho\in\cL(\cH)$,
 and 
 \beql{MO-5}
 \Pi(x)=\sum_{j}M_{xj}^{\da}M_{xj},
 \eeq
 where $\Pi$ is the POVM associated with the instrument $\cI$ given by \Eq{Inst-POVM}. i.e., $\Pi(x)=\cI(x)^{*}I$
 for all $x\in\R$, and conversely that every family of operators $\{M_{xj}\}$ satisfying
 \beql{MO-3}
 \sum_{xj}M_{xj}^{\da}M_{xj}=I
 \eeq
defines a completely positive instrument satisfying Eqs.~\eq{MO-4}, \eq{MO-5}. 
We call any family of operators $\{M_{xj}\}$ a family of {\em  measurement operators} if \Eq{MO-3} is satisfied.
Using measurement operators the General measurement axiom can be stated as follows. 
\bigskip

{\bf Axiom Q5' (General measurement axiom).}
{\em Every physically realizable apparatus $\bA(\bx)$ for the system $\bS$ with the
state space $\cH$ has a family of measurement operators $\{M_{xj}\}$  
such that the statistical properties of $\bA(\bx)$ are determined by 
\beqa
\Pr\{\bx=x\|\rh\}&=& \sum_{j}\Tr[M_{xj}^{\da}M_{xj}\rh], \\
\rh_{\{\bx=x\}}&=&\frac{\sum_{j}M_{xj} \rho M_{xj}^{\da}}{\sum_{j} \Tr[M_{xj}^{\da}M_{xj}\rh]},\quad
\mb{if \ $ \Pr\{\bx=x\|\rh\}>0$}
\eeqa
for all $x\in\R$ and $\rh\in\cS(\cH)$.  Conversely, every family of measurement operators $\{M_{xj}\}$ 
has at least one physically realizable apparatus $\bA(\bx)$ with the above statistical properties.
}
\bigskip

By the General measurement axiom, 
we can generalize the Wigner formula to arbitrary sequence of measurements,
and determine the joint probability distribution of the outcomes of
 any sequence of  measurements using physically realizable apparatuses.
 
\begin{Theorem}[Generalized Wigner formula]
Let $\cI_{1},\ldots,\cI_{n}$ be completely positive 
instruments for the system with
the state space $\cH$  in a state $\rh$ at time 0.
If one carries out measurements described by $\cI_{1},\ldots,\cI_{n}$  
at times $(0<)t_{1}<\cdots<t_{n}$ and otherwise leaves the system $\bS$ isolated
with the Hamiltonian $H$,
then the joint probability distribution of the outcomes 
$\bx_{1},\ldots,\bx_{n}$ of those measurements is given by
\beqa
\lefteqn{\Pr\{\bx_{1}=x_1,\bx_{2}=x_2,\ldots,\bx_{n}=x_n\|\rh\}}
\qquad\qquad\nn\\
&=&
\Tr[\cI_{n}(x_{n})\al(t_{n}-t_{n-1})\cdots
\cI_{2}(x_2)\al(t_{2}-t_{1})\cI_{1}(x_1)\al(t_{1})\rh],
\label{eq:GWF}
\eeqa
for any $x_1,x_2,\ldots,x_n\in\R$ and $\rh\in\cS(\cH)$,
where $\al$ is defined by 
$\al(t)\rh=e^{-iHt/\hslash}\rh e^{iHt/\hslash}$ for all 
$t\in\R$ and $\rh\in\cS(\cH)$. 
\end{Theorem}

Note that if the completely positive instruments  $\cI_{1},\ldots,\cI_{n}$ in the 
Generalized Wigner formula have the families of 
measurement operators $
\{
M^{(1)}_{{x_1}{j^{(1)}}}
\},\ldots,
\{
M^{(n)}_{{x_n}{j^{(n)}}}
\}$ the Generalized Wigner formula \Eq{GWF}
can be rewritten as
\beqa
\lefteqn{\Pr\{\bx_{1}=x_1,\bx_{2}=x_2,\ldots,\bx_{n}=x_n\|\rh\}}
%\qquad
\qquad\nn\\
&=&
\sum_{{j^{(1)},\ldots,j^{(n)}}}
\Tr[
M^{(n)}_{{x_n}{j^{(n)}}}
e^{-iH(t_{n}-t_{n-1})/\hslash}
\cdots
 M^{(2)}_{{x_2}{j^{(2)}}}
e^{-iH(t_{2}-t_{1})/\hslash}
 M^{(1)}_{{x_1}{j^{(1)}}}
    e^{-iHt_{1}/\hslash}\rh \nn\\
    & &\quad
         e^{iHt_{1}/\hslash}
(M^{(1)}_{{x_1}{j^{(1)}}})^{\da}
%\nn\\& &\quad
e^{iH(t_{2}-t_{1})/\hslash}
(M^{(2)}_{{x_2}{j^{(2)}}})^{\da}
\cdots
e^{iH(t_{n}-t_{n-1})/\hslash}
( M^{(n)}_{{x_n}{j^{(n)}}})^{\da}
].
\eeqa

Foundations of quantum measurement theory
based on the notion of completely positive instruments and indirect measurement models 
have been developed in  
\cite{83CR,84QC,85CA,85CC,86IQ,88MR,88MS,89RS,90QP,91QU,93CA,95MM,97QQ,97OQ,%
98QS,00MN,01OD,04URN,05PCN,06QPC,19A1}.

\end{document}